\documentclass[useAMS,usenatbib]{mn2e}
\usepackage{epsfig,graphics,graphicx,bm,amssymb}
\newcommand{\be}{\begin{equation}}
\newcommand{\ee}{\end{equation}}
\newcommand{\bea}{\begin{eqnarray}}
\newcommand{\eea}{\end{eqnarray}}
\newcommand{\bdm}{\begin{displaymath}}
\newcommand{\edm}{\end{displaymath}}

\begin{document} 

\title[MAMA: Multi-particle Algebraic Map for Accretion.]{MAMA: An Algebraic Map for the Secular Dynamics of Planetesimals
in Tight Binary Systems.}
\author[A.M. Leiva, J.A. Correa-Otto and C. Beaug\'e]
{A.M. Leiva$^*$$^{(1)}$, J.A. Correa-Otto$^{(1,2)}$ and C. Beaug\'e$^{(1,3)}$ \\
$(1)$ Observatorio Astron\'omico, Universidad Nacional de C\'ordoba, Laprida 854, (X5000BGR) C\'ordoba, Argentina \\
$(2)$ Instituto de Astronomia, Geof\'isica e Ci\^encias Atmosf\'ericas, USP, Rua do Mat\~ao 1226, 05508-900 S\~ao Paulo, Brazil \\
$(3)$ Instituto de Astronom\'ia Te\'orica y Experimental, Universidad Nacional de C\'ordoba, Laprida 854, (X5000BGR) C\'ordoba, Argentina \\
$^*$ e-mail: mleiva@oac.uncor.edu}

\maketitle

\begin{abstract}
We present an algebraic map (MAMA) for the dynamical and collisional evolution of a planetesimal swarm orbiting the main star of a tight binary system (TBS). The orbital evolution of each planetesimal is dictated by the secular perturbations of the secondary star and gas drag due to interactions with a protoplanetary disk. The gas disk is assumed eccentric with a constant precession rate. Gravitational interactions between the planetesimals are ignored. All bodies are assumed coplanar. A comparison with full N-body simulations shows that the map is of the order of $10^2$ times faster, while preserving all the main characteristics of the full system. 

In a second part of the work, we apply MAMA to the $\gamma$-Cephei, searching for friendly scenarios that may explain the formation of the giant planet detected in this system. For low-mass protoplanetary disks, we find that a low-eccentricity static disk aligned with the binary yields
impact velocities between planetesimals below the disruption threshold. All other scenarios appear hostile to planetary formation.
\end{abstract}


\section{Introduction}

Currently, there are more than 50 exoplanets detected in stellar binary systems (Chauvin et al. 2011). If the separation between the stellar components is larger than $\sim 50$ AU, the gravitational effects of the secondary star on a planetesimal or gas disk around the main star are small, and planetary formation is expected to proceed like in single stars. However, for compact (or tight) binary systems (hereafter, TBS), accretion can be seriously affected by the gravitational perturbations of the companion. Nevertheless, as many as 5 exoplanets are known to orbit individual components of TBS, the most extreme case being $\gamma$-Cephei, where the pericentric distance between the stellar components is only $\sim 12$ AU. 

Many dynamical and collisional studies may be found in the literature trying to understand the process of planetary formation in TBS (e.g. Marzari $\&$ Scholl 2000, Th\'ebault et al. 2004 y 2006, Paardekooper et al. 2008, Beaug\'e et al. 2010, Th\'ebault 2011). So far, all attempts have been unsuccessful. The gravitational perturbations of the secondary star are too large and systematically lead to impact velocities beyond the disruption limit. Recently, Rafikov (2013) showed that the gravitational interaction with the gas disk could counteract the effects of the binary, significantly reducing the collisional velocities of the swarm to acceptable levels. However, it appears that this requires a very massive disks, of the order of $0.1 M_\odot$. Since another effect of the binary is a severe truncation and mass loss of the original protoplanetary disk (e.g. Artymowicz \& Lubow 1994), it is not clear whether such massive disks would be expected in these systems.

Paardekooper et al. (2008) discussed that low collisional velocities could, in principle, be attained if the gas disk was permanently aligned with the binary and had an eccentricity similar to the forced eccentricities of the planetesimals. However, since at that time hydro-simulations showed precessing disks, this idea was not pursued. Recent years have shown a variety of hydrodynamical simulations of the dynamics of circumstellar gas disks in TBS (e.g. Kley \& Nelson 2008, Kley et al. 2008, Marzari et al. 2009, Marzari et al. 2012, M\"uller \& Kley 2012), adopting different thermodynamic properties and boundary conditions. While isothermic massless disks show moderate-to-high eccentricities ($e_g \sim 0.1-0.2$) and relatively high retrograde precession rates ($|{\dot \varpi_g}| = |g_g| \sim 2\pi/1000$ yr$^{-1}$), radiative disks and self-gravity seem to favor more circular and static disks (e.g. Marzari et al. 2012, M\"uller \& Kley 2012). However, the results seem very sensitive to disk parameters, including the initial disk aspect ratio $H/r$ and $\alpha$-viscosity.

\begin{figure}
\begin{center}
\epsfig{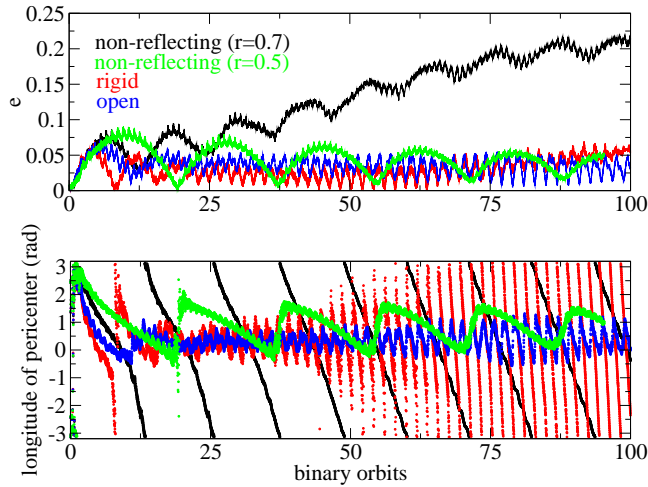}
\caption{FARGO simulations of a circumstellar disk around $\gamma$-Cephei-A. Plots show the averaged gas eccentricity ($e_g$) and longitude of the pericenter $\varpi_g$ as a function of time, for four different inner boundary conditions. For the first two runs, $r$ is the inner radius of the disk.}
\label{fig1}
\end{center}
\end{figure}

All previous simulations employed grid (i.e. Eulerian) codes, such as FARGO (Masset 2000) or RH2D (Kley 1999). Although Eulerian methods have proved very reliable for disks around single stars, there are some indications that they may be problematic for TBS. Figure \ref{fig1} shows four different FARGO simulations of a gas disk around $\gamma$-Cephei-A, perturbed by its binary companion. Mases and orbital elements we chosen following the best radial velocity fit by Hatzes et al. (2003). We adopted an initial $r^{-1/2}$ surface density profile with $\Sigma(r=1) = 7 \times 10^{-4}$ gr/cm$^2$ an $\alpha$-viscosity of $\alpha = 10^{-5}$. In all cases we chose an open outer boundary condition, but changed the inner boundary condition, as shown in the top left-hand corner of the upper plot. The resulting dynamics of the gas disk is very different, even far from the inner edge. In some cases the disk precesses, while in others the disk appears static. The behavior of the eccentricity is also sensitive to the boundary condition, although perhaps in a lesser extent. Similar results were also found by Kley et al. (2008). 

A different problem is related to the timescale of the simulations. All predictions of the dynamics of protoplanetary disks are extrapolated from just $\sim 10^2$ orbital periods of the binary, even though many cases show evidence that the system has not yeat reached an equilibrium. It may occur that secular perturbations from the binary would modify the results of the simulations, and these would only be noticeable on the long run. Since we expect planetary formation to take at least $\sim 10^5$ orbital periods of the binary, we wonder whether what we see in the short term is necessarily indicative of the long term behavior.

From these considerations, we believe that the real long-term equilibrium configuration of circumstellar disks in TBS is far from established. So, instead of adopting a given recipe for the gas dynamics and proving (or disproving) planetary formation in such a scenario, for the present paper we have chosen the inverse route. We will take the gas eccentricity and precession rate as variables in a parameter space, and search for those values that allow constructive collisions of a planetesimal swarm around the primary star of a TBS. Since this route implies a larger series of numerical simulations, we will present an algebraic map (dubbed MAMA) that facilitates this analysis. In particular, we will apply our map to $\gamma$-Cephei, a well known and amply discussed system. Our results could guide future studies in disk dynamics trying to discover what thermodynamics properties they entail, or, conversely, if other planetary formation scenarios are required. 

This manuscript is divided as follows: In section 2, we review the differential equations governing the dynamics of small planetesimals affected by the gravitational perturbation of the stellar companion and the drag force from the gas. In Section 3 we construct the algebraic map (MAMA) for TBS, and compare its performance with respect to full N-body simulations. The application of MAMA to the $\gamma$-Cephei system is discussed in Section 4, where we search for disk parameters leading to accretion-friendly scenarios. Finally, conclusions close this work in Section 5.

\section{Dynamics evolution of small planetesimals in TBS}

We begin assuming a tight binary system (TBS) composed of a main star of mass $m_{\rm A}$ and a stellar companion of mass $m_{\rm B}$. We choose a coordinate system centered in $m_{\rm A}$ with the $z$-axis parallel to the orbital angular momentum of the system. In this reference frame, we will denote by $a_{\rm B}$ the semimajor axis of the secondary, $e_{\rm B}$ its eccentricity and $\varpi_{\rm B}$ its longitude of pericenter (the origin of all longitudes is arbitrary).

We also assume that both the gas disk and the planetesimal disk orbit the primary star in the same plane. All orbital elements be $m_{\rm A}$-centric. Our focus will then be on the dynamics of planetesimals when subject to gas drag and the gravitational perturbation of the secondary star.

A full study of this dynamics is a complicated task; however, it may be simplified considering a linear superposition of two interactions: {\it (i)} the drag gas with the disk (Weidenschilling et al. 1997, Supulver \& Lin 2000, Beaug\'e et al. 2010), and {\it (ii)} the gravitational perturbation of the secondary (Heppenheimer 1978, Th\'ebault et al. 2006, Giuppone et al. 2011). Each is discussed in the following sub-sections.

\subsection{Gas drag}

For spheric planetesimals with radius $s > 0.1$ km, the gas drag is a non-linear function of the relative velocity ($v_{\rm rel}$) with respect to the gas, and its magnitude is proportional to $v_{\rm rel}^2$ (Adachi et al. 1976, Weidenschilling et al. 1997, Supulver \& Lin 2000). The acceleration suffered by the planetesimal is given by
\be
\label{eq1}
{\ddot {\bf r}} = -{\cal C} |{\bf v}_{\rm rel}| {\bf v}_{\rm rel},
\ee
where
\be
\label{eq2}
{\cal C} = \frac{3 C_D}{8} \frac{1}{s} \frac{\rho_{\rm g}}{\rho_{\rm p}}.
\ee
Here $\rho_{\rm p}$ and $\rho_{\rm g}$ are the volume densities of the planetesimal and gas, respectively, and $C_D=0.44$ is an adimensional drag coefficient, usually considered constant for high Reynold numbers (Weidenschilling et al. 1997). 

In a tight binary system, the gravitational perturbations of $m_{\rm B}$ cause drastic changes in the surface density of the gas disk. According to hydro-simulations (Paardekooper et al. 2008, Kley \& Nelson 2008), the resulting surface density profile is almost linear up to an outer limit $a_{\rm out}$, whose value is close to the location of the $L_1$ Lagrange point. Following Beaug\'e et al. (2010), we adopt a functional form for $\rho_{\rm g}$ given by
\be
\label{eq3}
{\rho_{\rm g}}(a)\simeq \frac{3}{2 \pi} \frac{M_{\rm T}}{ a_{\rm out}^{3}H_{\rm R}} \biggl( \frac{a_{\rm out}}{a} - 1 \biggr),
\ee
where $M_T$ is the total mass of the disk, $H_R=0.05$ is its scale height, and $a$ the semimajor axis of each gas element. For $\gamma$-Cephei, we find that $a_{\rm out} \simeq 5$ AU.

The gas disk has a negative pressure gradient which causes it to orbit $m_{\rm A}$ with a sub-Keplerian velocity: ${\bf v}_g = \alpha {\bf v}_{\rm Kep}$. Following Adachi et al. (1976), we assume $\alpha=0.995$. Then, the relative velocity between a planetesimal and a gas element, both at a given position {\bf r}, is given in polar coordinates by ${\bf v}_{\rm rel}=v_r \hat{{ r}} + v_{\theta} \hat{{ \theta}}$, where
\bea
\label{eq4}
v_r &=&  \sqrt{ \frac{\mu}{p} } \biggl[  e \sin{(f)} - \alpha \ e_{\rm g}  \sin{(f+\Delta \varpi)} \cdot \nonumber \\
    & & \;\;\;\;\;\;\; \cdot \biggl( \frac{1+e \cos{(f)}}{1+e_{\rm g} \cos{(f+\Delta \varpi)}}\biggr)^{\frac{1}{2}}\biggr] \\
v_{\theta} &=&  \sqrt{ \frac{\mu}{p} } \; \biggl[  (1 + e \cos{f}) - \alpha \ (1 + e_{\rm g}  \cos{(f+\Delta \varpi)}) \cdot \nonumber \\
    & & \;\ \ \ \ \ \cdot \biggl( \frac{1+e \cos{(f)}}{1+e_{\rm g} \cos{(f+\Delta \varpi)}}\biggr)^{\frac{1}{2}}\biggr] \nonumber .
\eea
In these expressions $a$, $e$, $\varpi$ and $f$ are the semimajor axis, eccentricity, longitude of pericenter and true anomaly of the planetesimal, $\Delta \varpi=\varpi-\varpi_{\rm g}$, $\mu = {\cal G} m_{\rm A}$, ${\cal G}$ the gravitational constant, and $p=a(1-e^{2})$ the semi-lactus rectum. The reader is referred to Beaug\'e et al. (2010) for more details.

The variational equations for the reduced set of variables ($a$, $e$, $\varpi$) can be obtained from Gauss' perturbation equations (e.g. Roy 2005):
\bea
\label{eq5}
\frac{da}{dt}\biggr|_{\rm drag} &=& \frac{2a^2}{\sqrt{\mu p}} 
\biggl( R' \; e \sin{f} +  T' \; (1+e \cos{f}) \biggr) \nonumber \\
\frac{dk}{dt}\biggr|_{\rm drag} &=& \sqrt{\frac{p}{\mu}}  \biggl( R' \; \sin{(f+\varpi)} \ +
\\
 & & \ \ \ \ \ + \ T' \; \frac{(2+e \cos{f}) \cos{(f+\varpi)}+ e \cos{\varpi}}{1+e \cos{f}}  \biggr)  \nonumber \\
\frac{dh}{dt}\biggr|_{\rm drag} &=& \sqrt{\frac{p}{\mu}}  \biggl( -R' \; \cos{(f+\varpi)} \ + 
\nonumber \\
 & & \ \ \ \ \ + \ T' \; \frac{(2+e \cos{f}) \sin{(f+\varpi)}+ e \sin{\varpi}}{1+e \cos{f}}  \biggr) \nonumber
\eea
where $(k,h)=(e \cos \varpi,e \sin \varpi)$ are the Cartesian analogues of $(e,\varpi)$. Functions $R'$ and $T'$ are the radial and transverse component of the acceleration due to the gas drag and are defined by $R' = -{\cal C} |{\bf v}_{\rm rel}| \ v_r$ and $T' = -{\cal C} |{\bf v}_{\rm rel}| 
\ v_{\theta}$.

\subsection{Secular gravitational perturbations}

Our study is performed in the restricted three-body problem. This implies that we will disregard the perturbations of the planetesimals and gas on $m_{\rm B}$, which will move in a fixed elliptical orbit around the main star. Also, we will neglect the mutual gravitational interactions between the planetesimals themselves. 

In this scenario, and outside any significant mean-motion resonances between the planetesimals $m$ and $m_{\rm B}$, the gravitational dynamics of the small bodies will be dominated by secular perturbations, as well as short-period terms associated to the mean longitudes. These latter contributions can be eliminated by a perturbation technique known as averaging, in which the osculating variables $(a,k,h,f)$ and transformed to averaged variables $(a^*,k^*,h^*,f^*)$ which do not contain the short-period variations. In the averaged (secular) system, the ``proper'' semimajor axis $a^*$ is constant and the only pertinent variables are $(k^*,h^*)$. The resulting equations of motion are then
\be
\label{eq6}
\frac{dk^*}{dt} = - g h^* \hspace*{0.5cm} ; \hspace*{0.5cm} \frac{dh^*}{dt} = g (k^* - e_{\rm f}),
\ee
where $g$ is the secular frequency and $e_f$ is the forced eccentricity. Using a second-order Hori-type averaging procedure, Giuppone et al. (2011) found approximate expressions for both quantities:
\bea 
\label{eq7}
g &=& \frac{3}{4} \frac{m_{\rm B}}{m_{\rm A}} \frac{n^* {a^*}^3}{a_{\rm B}^3(1-e_{\rm B}^2)^{3/2}}
      \left[ 1+32 \left( \frac{m_{\rm B}}{m_{\rm A}} \right) \left( \frac{a^*}{a_{\rm B}} \right)^2 (1-e_{\rm B}^{2})^{-5} \right], \nonumber \\
e_{\rm f} &=&  \frac{5}{4} \frac{a^*}{a_{\rm B}} \frac{e_{\rm B}}{(1-e_{\rm B}^2)} \left[ 1-16 \left( \frac{m_{\rm B}}{m_{\rm A}} \right) \left( \frac{a^*}{a_{\rm B}} \right)^2 (1-e_{\rm B}^{2})^{-5} \right] ,
\eea
where $n^*$ is the proper mean motion. The terms within the square brackets are the second-order contributions and do not appear in first-order theories such as Heppenheimer (1978). The reader is referred to Giuppone et al. (2011) for a comparison between both secular models. 

The secular system (\ref{eq6}) is linear and can be easily solved analytically. Given initial conditions  $(a_0^*,k_0^*,h_0^*)$, we can write
\bea
\label{eq8}
a^*(t)&=& a_0^* \nonumber\\
k^*(t)&=& e_{\rm p} \cos{(g \ (t-t_0)+\phi_0)}+e_{\rm f}  \\
h^*(t)&=& e_{\rm p} \sin{(g \ (t-t_0)+\phi_0)} \nonumber,
\eea
where $t_0$ is the initial time, $e_{\rm p}^{2}=(k_0^*-e_{\rm f})^{2}+(h_0^*)^{2}$ and $\tan \phi_0=h_0^*/(k_0^*-e_{\rm f})$. The quantity $e_{\rm p}$ is sometimes referred to as the free eccentricity.

\subsection{Linear superposition of the two interactions}

Our complete model will be the direct sum of equations (\ref{eq5}) and (\ref{eq6}). However, it must be noted that the secular model (\ref{eq6}) was constructed with the averaged orbital elements while the drag model (\ref{eq5}) assumes osculating elements. Even so, since $a$ exhibits periodic variations around $a^*$, a more precise reproduction of the orbital decay from gas drag will be obtained if we adopt $a^*$ instead of $a$ in equations (\ref{eq5}).

To merge both sets of differential equations, we must to find a relation between osculating and proper variables. Instead of employing cumbersome canonical transformations (e.g. Giuppone et al. 2011), in this paper we preferred a purely numerical approach.

Preliminary tests showed that the difference between $(k^*,h^*)$ and $(k,h)$ is not significant to the overall evolution of the system. Since our aim is to keep the complete model as simple as possible, we opted for neglecting the transformation of the secular variables. As we will show below, this approximation is good enough to our purposes. The difference between $a$ and $a^*$, on the other hand, are mainly noticeable in the orbital decay timescale. Although the errors introduced by neglecting the transformation $a \rightarrow a^*$ are not large (of the order of $\sim 1-5 \%$), they are easily remedied. 

The functional form $a^*(a)$ was built numerically. First, we performed N-body simulations for the dynamical evolution of the semimajor axis of several test particles, each with a different initial value $a_0 \in [1,5]$ AU, $e_0=0$  and mean anomaly $M_0=0^\circ$. The initial value of $e$ is not important, since the most important term in the amplitude of $a$ is of order zero in the eccentricity. 

\begin{figure}
\begin{center}
\epsfig{figure=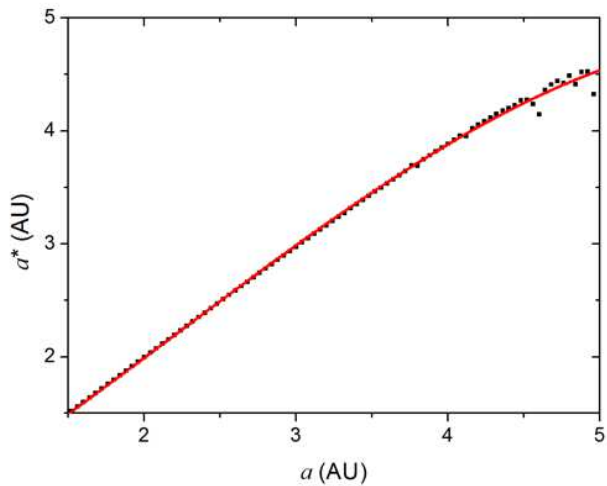,width=8.5cm,angle=0,clip=}
\caption{Relation between the osculating ($a$) and proper ($a^*$) semimajor axis, the black dots show  numerical results, while the red curve corresponds to the empirical fit (\ref{eq9}).}
\label{fig2}
\end{center}
\end{figure}

In all cases we adopted the $\gamma$-Cephei binary system. The output $a(t)$ of each simulation was then transformed to $a^*$ using a low-pass FIR (finite impulse response) filter (e.g. Carpino et al. 1987) designed to remove all periodic variations up to 5 orbital periods of the binary. The resulting distribution of $a^*$ for each initial osculating $a_0$ is plotted in Figure \ref{fig2} (black circles). The red curve shows a numerical fit using a cubic polynomial in $a$, whose expression is:
\be 
\label{eq9}
a^*=0.21959 + 0.67350 a + 0.14975 a^2 - 0.02237 a^3 ,
\ee 
The agreement with the numerical results is very good. However, it is important to keep in mind that this polynomial is only valid for initial conditions with $M_0=0$. Thus, in all numerical simulations performed in this work we will adopt the same initial mean anomaly for the particles. 

With the empirical relation (\ref{eq9}) between the osculating and proper semimajor axis, we can construct the complete model. Then, with the linear superposition of the models (\ref{eq5}) and (\ref{eq6}) we obtain the complete dynamical model as:
\bea
\label{eq10}
\frac{da}{dt}  &=&  \frac{da}{dt}\big|_{\rm drag} \nonumber \\
\frac{dk}{dt}  &=&  \frac{dk}{dt}\big|_{\rm drag}-g \ h \\
\frac{dh}{dt}  &=&  \frac{dh}{dt}\big|_{\rm drag}+g \ (k-e_{\rm f})  \nonumber
\eea
where the drag terms in the two latter equations must also be evaluated at $a$ using its relationship with $a^*$.  

\subsection{Comparison between N-body simulations and the secular model}\label{sec:nada}

To test the accuracy of the model, we chose once again the $\gamma$-Cephei system as our working example. We first performed a series of N-body simulations of the evolution of the planetesimals with different physical radii and initial conditions, and then compared the results with numerical integrations of equations (\ref{eq10}). In both cases the differential equations were solved with a Bulirsch-Stoer code using an accuracy of $10^{-11}$.

\begin{figure}
\begin{center}
\epsfig{figure=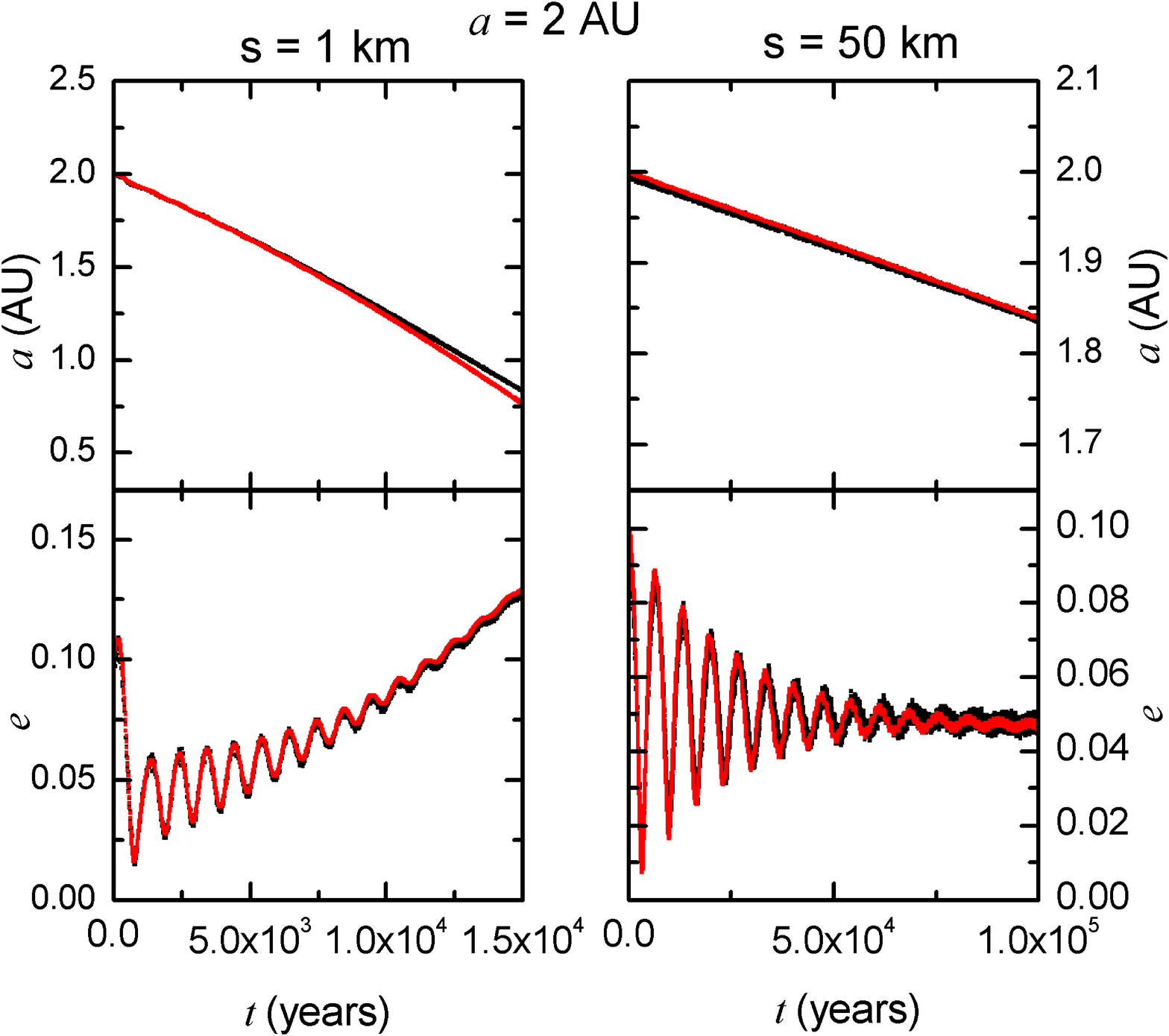,width=9cm,angle=0,clip=}
\caption{Temporal evolution of $a$ and $e$ for two different size planetesimals, $s=1$ km (left) and $s=50$ km (right). Initial osculating semimajor axis was chosen equal to $a_0 = 2$ AU. The black dots show the results of the full N-body numerical simulation, while the red curves correspond to model (\ref{eq10}).}
\label{fig3}
\end{center}
\end{figure}

\begin{figure}
\begin{center}
\epsfig{figure=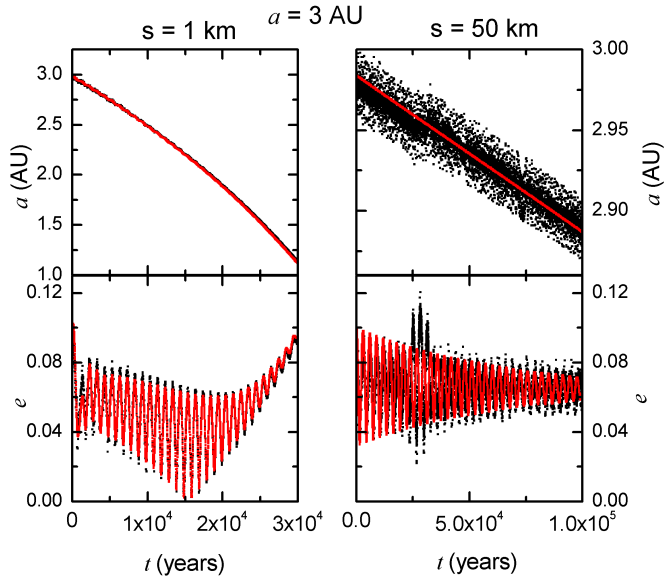,width=9cm,angle=0,clip=}
\caption{Same as previous figure, but now the initial osculating semimajor axis was $a_0=3$ AU.}
\label{fig4}
\end{center}
\end{figure}

Figures \ref{fig3} and \ref{fig4} show two extremes cases. Plots on the left correspond to planetesimals with radii $s=1$ km, while those on the right to $s=50$ km, both assuming $\rho_{\rm p} = 3$ gr/cm$^3$. In Figure \ref{fig3} the initial osculating semimajor axis was chosen equal to $a_0=2$ AU, while in Figure \ref{fig4} this value was increased to $a_0=3$ AU. Other orbital elements were $M=0^\circ$, $e_0=0.1$ and $\varpi_0=0^{\circ}$. We assumed an eccentric gas disk ($e_{\rm g}=0.2$) with a retrograde precession rate equal to $2\pi/|g_{\rm g}|=1000$ years. The disk was further assumed to have a volume density of $\rho_{\rm g} (2$AU$) =5 \times 10^{-10}$ gr/cm$^3$ (Paardekooper et al. 2008, Beaug\'e et al. 2010) and an outer truncation radius $a_{out}=5$ AU. 

From these results we can see that the dynamical behavior of small planetesimals ($s=1$ km, left panels) is well reproduced by our model. For these bodies, the interaction with the gas disk is dominant over the gravitational perturbations from the binary. The simple averaged equations give a correct qualitative (and quantitative) prediction about the orbital decay, as well as the amplitude, frequency and damping of the eccentricity.

For large planetesimals ($s=50$ km, right panels), on the other hand, the gravitational perturbations from $m_{\rm B}$ are more important than the drag gas. For the planetesimal of initial semimajor axis $a_0 = 2$ AU (Fig. \ref{fig3}), the model gives a very good approximation to the real dynamics. The same agreement is also observed for larger initial semimajor axis ($a_0=3$ AU, Fig. \ref{fig4}), although our model fails to reproduce an excitation in the eccentricity and a temporary jump in the semimajor axis, both occurring simultaneously at $t \sim 2.5 \times 10^4$ years. A closer look reveals that this behavior is generated by a passage through a high-order mean-motion resonance (MMR) with the binary (see Giuppone et al. 2011 for more detailed examples). Since resonant interactions are not included in our model, equations (\ref{eq10}) are unable to reproduce this effect. Nevertheless, with the exception of the resonance scattering, the results of the model seem very accurate.

\section{High-Order Mean-Motion Resonances in the $\gamma$-Cephei System}

Since the outer parts of both the gas and planetesimal disks may lie in regions affected by MMR, they could seriously impair the use of our secular model. Thus, before proceeding in the construction of our algebraic map, it is important to evaluate the effects of high-order commensurabilities in the possible accretion process. Our first analysis along these lines will be to map the regions of regular and chaotic motion for a wide range of initial conditions. As before, we adopt $\gamma$-Cephei as our working example.

We considered a grid of $3000 \times 201$ initial conditions in the semimajor axis vs. eccentricity plane, with values in the intervals $a \in [2,5]$ AU and $e \in [0.0,0.2]$. The number of points in each axis correspond to an equal step of $\Delta a = 0.001$ AU and $\Delta e = 0.001$ between successive points. Starting values for the angles where taken equal to zero, except for $\lambda=M+\varpi$ which has taken equal to $180^\circ$. All the test particles were integrated for $2 \times 10^5$ years (equal to $\sim 3500$ orbital periods of the binary) using an N-body code with a Bulirsch-Stoer integrator (precision $ll=10^{-12}$).

For each orbit we calculated the averaged MEGNO indicator $\langle Y \rangle$ (Cincotta \& Sim\'o 2000). This quantity has proven to be an efficient identifier of chaotic behavior, been significantly faster than the classical maximum Lyapunov exponent. Recall that values $\langle Y \rangle \le 2$ correspond to regular orbits, while $\langle Y \rangle > 2$ are indicative of chaotic motion.

\begin{figure}
\begin{center}
\epsfig{figure=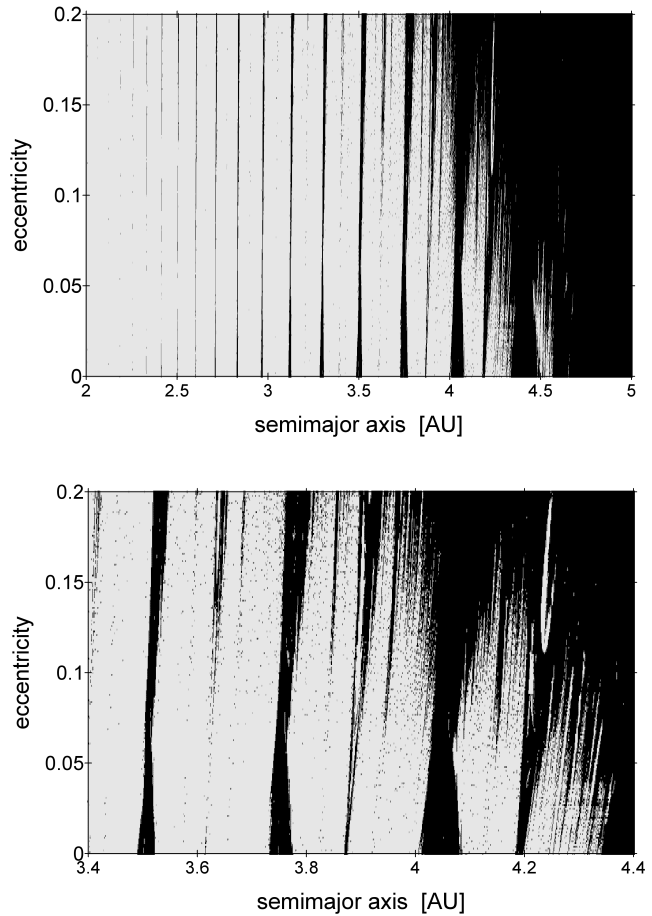,width=8.5cm,angle=0,clip=}
\caption{MEGNO map of $3000 \times 201$ initial conditions in the $a$-$e$ plane, each corresponding to test-particles in the $\gamma$-Cephei binary system. Total integration time was equal to $2 \times 10^5$ years. Light gray dots indicate regular orbits, while black dots correspond to chaotic solutions.}
\label{fig5}
\end{center}
\end{figure}

Results are shown in Figure \ref{fig5}, where the color associated to each initial conditions is related to the final value of $\langle Y \rangle$. Regular orbits are shown in light gray, while chaotic solutions are shown in black. The top plot presents the complete map, while the bottom graph zooms in on the interval between $3.4$ and $4.4$ AU.

For $a < 4$ AU, most of the phase plane is dominated by regular orbits, crossed by thin almost-vertical stripes of chaotic motion, each associated to a different MMR. This far from the perturber, the resonances are isolated and their effect is restricted to a small region around their center. Conversely, the outer region of the map beyond $4$ AU is almost completely chaotic, with only small areas of regular motion at low eccentricities. In this region the libration width of the MMRs are sufficiently large to allow overlap even for moderate eccentricities and, thus, cause the appearance of zones characterized by global chaos. Finally, for $a > 4.5$ AU the resonance overlap is complete even for quasi-circular orbits, and all initial conditions are dynamically unstable. Notice how the outer limit of the gas disk in this system (located near $5$ AU, according to hydrodynamical simulations) shows a good agreement with the region of the phase plane immersed in a chaotic sea even for circular orbits. 

To estimate which MMR are associated to each chaotic zone, we can calculate its position from Kepler's third law. Given a generic $(p+q)/p$ commensurability, the nominal (i.e. exact) resonant semimajor axis is given by
\begin{equation}
\label{eq11}
a = a_B \; \biggl( \frac{m_A}{m_A+m_B} \biggr)^{1/3} \biggl( \frac{p}{p+q}
\biggr)^{2/3} .
\end{equation} 
Table \ref{tab1} shows the nominal position of several first-degree resonances in the region of interest. Here we have adopted the classical nomenclature in which the value of $q$ gives the {\it order} of the commensurability, while $p$ is its {\it degree}. A comparison between these positions and the dynamics maps in Figure \ref{fig5} shows two important results.

First, all the stripes of strong chaotic motion in the top panel are associated to first-degree MMR of high-order. In Solar System problems, high-order resonances have negligible dynamical consequences, but in the present system the combination of a large mass and high eccentricity perturber enhances their effects. Later on we will analyze just how important they can be in the dynamical evolution of test planetesimals. Thinner stripes, specially noticeable in the lower plot, correspond to second and third-degree resonances. The region $a > 4$ AU is specially rich in these structures forming a forest of lines that contribute to the resonance overlap.

A second result is that the locations of the resonances are significantly shifted with respect to the exact semimajor axes. This is expected from what is sometimes known as the ``Law of Structure'' (Ferraz-Mello, 1988) or the ``pericentric branch'' (e.g. Moons \& Morbidelli 1993). Basically, this means that the center of the resonance domain is a function of the eccentricity, and is usually shifted away from the exact semimajor axis. The magnitude of this shift is very sensitive with respect to the system parameters, particularly $m_B$ and $e_B$.

\begin{table}
\begin{center}
\begin{tabular}{cc}
\hline
 $(p+q)/p$ & $a$ [AU]    \\ 
\hline
 $\;$8/1 & 4.300 \\ 
 $\;$9/1 & 3.970 \\
 10/1 & 3.700 \\
 11/1 & 3.472 \\
 12/1 & 3.276 \\
 13/1 & 3.106 \\
 14/1 & 2.956 \\
 15/1 & 2.823 \\
 16/1 & 2.705 \\
 17/1 & 2.597 \\
 18/1 & 2.500 \\
\hline
\end{tabular}
\end{center}
\caption{Nominal semimajor axis of several first-degree MMR in the $\gamma$-Cephei system.}
\label{tab1}
\end{table}

\begin{figure}
\begin{center}
\epsfig{figure=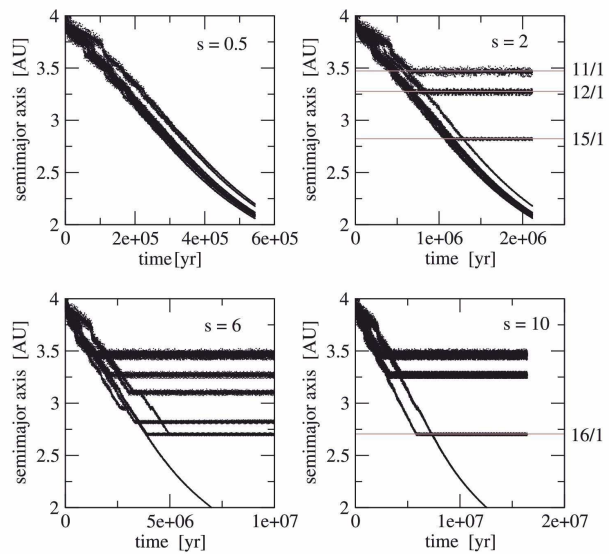,width=8.5cm,angle=0,clip=}
\caption{N-body simulations of planetesimals in the $\gamma$-Cephei system, under the combined effects of gravitational perturbations and gas drag. Each panel shows the semimajor axis, as function of time, of a set of $10$ fictitious bodies with $a_0=4$ AU, $e_0=0.1$ and random initial mean anomalies. The radii $s$ of the planetesimals are indicated (in km) in the top right-hand corner of each plot. The location of the strongest MMR are shown in both right-hand graphs.}
\label{fig6}
\end{center}
\end{figure}

While Figure \ref{fig5} appears to indicate limited effect of MMRs on planetesimal orbits with $a < 4$ AU, these dynamical maps correspond to the conservative problem in which the effects of the gas drag are not considered. Figure \ref{fig6}, on the other hand, shows the evolution of the semimajor axis of four different sets of $10$ fictitious planetesimals, again in the $\gamma$-Cephei system, with a non-linear drag. Initial conditions were chosen equal to $a_0=4$ AU, $e_0=0.1$, $\Delta \varpi = \varpi - \varpi_B = 0$, and random values of the mean anomaly $M_0$. The gas disk was assumed static (no precession) and with a small eccentricity ($e_g=0.05$). In the top left-hand panel we considered small planetesimals with physical radii $s=0.5$ km, while for the bottom right-hand plot we used $s=10$ km. Other plots correspond to intermediate values.

Giuppone et al. (2011) showed that some planetesimals could undergo resonance trapping, even though the effects of the drag leads to divergent migration. Here we can see the same effect in more detail, and how it varies according to the size of the planetesimal. For very small bodies the orbital decay is very pronounced and cannot be overcome by the resonant perturbations; consequently no resonance trapping is observed and all passages are characterized by temporary excitations of the eccentricity. In a little over $5 \times 10^5$ yrs all the planetesimals have already fallen below the semimajor axis of the observed planet (i.e. $\sim 2$ AU), and the overall dynamical evolution is primarily dictated by gad drag and secular gravitational effects.

For larger bodies, the timescale for orbital decay is longer than the typical libration period within the MMR. Resonance trapping is now possible, even though the commensurabilities are of high-order. For $s=2$ km, only $\sim 30\%$ of the bodies are trapped, while this number increases to about $\sim 90 \%$ for $s=10$ km. The $11/1$ MMR is the preferred location, although some trapping is also observed in other commensurabilities down to the $16/1$. However, for $s>6$ km the orbital decay towards the resonance already takes longer than the expected timespan of the gas disk, so it is questionable whether this effect would be dynamically significant in real systems. 

These simulations were performed for a static gas disk. As shown by Beaug\'e et al. (2010), a precessing disk causes a faster orbital decay, so resonance trapping is less effective in those cases. Simulations using a retrograde precession rate of $g_g = -2 \pi / 1000$ yr$^{-1}$ show that resonance trapping is ineffective for $s<50$ km. 

In conclusion, we have found that resonant effects should be important in the dynamical evolution of relatively large planetesimals with initial semimajor axis $a > 3$ AU, especially in static gas disks. In those cases our secular model should be used with caution. However, for initial conditions closer to the star, MMR seem to be of little consequence, and our model described by equations (\ref{eq10}) can constitute a working and adequate approximation to the real dynamics.

\section{The algebraic map MAMA}\label{secmama}

Although our mixed-secular model (\ref{eq10}) is much faster than a full N-body simulation of the exact equations, we can drastically improve its performance by the implementation of an algebraic map. This is desirable if we want to test many different system parameters, searching for the most friendly scenario for the process of accretion. 

Following the pioneering works of Malhotra (1994), Cordeiro et al. (1997) and Mikkola (1997), we can construct our algebraic map as an extension of the classical leap-frog algorithm for dissipative systems. 
We begin by rewriting the complete analytical model as:
\bea
\label{eq12}
\frac{da}{dt}  &=&  \frac{da}{dt}\big|_{\rm grav} + \frac{da}{dt}\big|_{\rm drag} \nonumber \\
\frac{dk}{dt}  &=&  \frac{dk}{dt}\big|_{\rm grav} + \frac{dk}{dt}\big|_{\rm drag} \\
\frac{dh}{dt}  &=&  \frac{dh}{dt}\big|_{\rm grav} + \frac{dh}{dt}\big|_{\rm drag}  \nonumber
\eea
where the first term of the r.h.s. is the gravitational contribution from the binary, for which
$\frac{da}{dt}\big|_{\rm grav} \equiv 0$. While the gravitational terms define an autonomous system, the drag terms include the time implicitly through the true anomaly $f$. 

Defining $\Delta t$ as the time-step, the algebraic map is constructed by the following sequence of steps:

\begin{itemize}
\item Step 1 (Drift): $(a^*_0,k_0,h_0,M_0) \rightarrow (a^*_1,k_1,h_1,M_1)$

Given initial conditions $(a^*_0,k_0,h_0,M_0)$, where $M_0$ is the mean anomaly at $t=t_0$, we integrate system (\ref{eq12}) considering only the conservative terms for half a time-step $\Delta t/2$. Since this "unperturbed" system has an analytical solution in closed form, we simply obtain:
\bea
\label{eq13}
a^*_1 &=& a^*_0 \nonumber \\
k_1 &=& e_{\rm p} \cos{(g \ \Delta t/2 + \phi_0)} + e_{\rm f} \\
h_1 &=& e_{\rm p} \sin{(g \ \Delta t/2 + \phi_0)} \nonumber,
\eea
where the values of $e_{\rm p}$, $e_{\rm f}$, and $g$ are calculated at $a^*_0$. The mean anomaly is estimated with $M_1=n^*_0 \ \Delta t/2 + M_0$. This is obviously an approximation, since we are considering the mean mean-motion $n^*_0$ instead of its osculating value, but test simulations (see Figures \ref{fig3} and \ref{fig4}) show the error is not significant.

\item Step 2 (Kick): $(a^*_1,k_1,h_1,M_1) \rightarrow (a^*_2,k_2,h_2,M_2)$

We now apply a first-order integration, applying solely the drag effects, for a time-step $\Delta t$
\bea
\label{eq14}
a^*_2 &=& a^*_1 + \Delta T \ \frac{da^*}{dt}\big|_{\rm drag} \nonumber\\
k_2   &=& k_1   + \Delta T \ \frac{dk}{dt}\big|_{\rm drag} \\
h_2   &=& h_1   + \Delta T \ \frac{dh}{dt}\big|_{\rm drag} \nonumber.
\eea
where the value of the true anomaly $f_1$ in the drag equations is determined from the mean anomaly $M_1$ solving Kepler's equation. The mean anomaly $M_2$ is left unchanged, so that $M_2=M_1$.

\vspace*{0.3cm}
\item Step 3 (Drift): $(a^*_2,k_2,h_2,M_2) \rightarrow (a^*_3,k_3,h_3,M_3)$

Finally, we repeat Step 1 for a time-step $\Delta t/2$, updating the initial conditions and values of $e_{\rm p}$, $e_{\rm f}$, and $g$ according to the new proper semimajor axis.
 
\end{itemize}

As usual, after the first application of the map, both drifts can be fused into a single application of the conservative equations for a full time-step interval $\Delta t$. This scheme defines our algebraic map for the complete model, hereafter referred to as MAMA.

\subsection{Step time for \textit{MAMA}}

In order to apply MAMA successfully, we must specify a value for the time-step $\Delta t$ that guarantees a fast code with accurate results. Once again, we considered the $\gamma$-Cephei system as an example, and assumed an elipitical disk ($e_g=0.1$) around the main star $m_{\rm A}$. We then analyzed the dynamical evolution of five different planetesimals ($s=1$, $5$, $10$, $20$ and $50$ km), comparing the full numerical solutions of the complete model (\ref{eq10}) with the application of the algebraic map. Both integration methods were followed for $10^3$ years, at the end of which we calculated the relative difference in semimajor axis and eccentricity (denoted by er$_A$ and er$_E$, respectively). 

\begin{figure}
\begin{center}
\epsfig{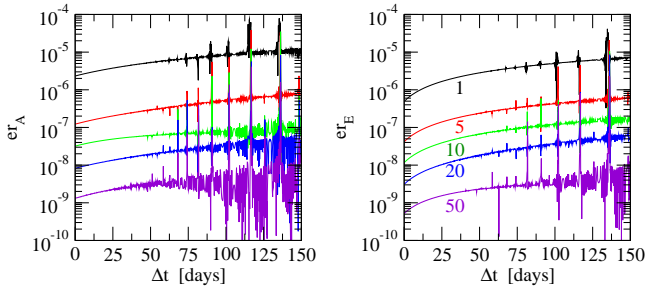}
\caption{Maximum relative errors for the semimajor axis ($er_A$) and the eccentricity ($er_E$) as function of the step time of MAMA, and for planetesimals of different size. Black: $s=1$, Red: $s=5$, Green: $s=10$, Blue: $s=20$, Violet: $s=50$, where all values are given in kilometers.}
\label{fig7}
\end{center}
\end{figure}

Results are shown in Figure \ref{fig7} for initial semimajor axis $a_0=2$ AU, where there is an evident increase in the error for smaller value of $s$, for which the drag term is more important. 
Although for small values of $\Delta t$ the error is linear with the step size (as expected from a leap-frog based map), we also note the appearance of localized peaks in the errors, that occur for the same values of $\Delta t$ independently of the particle size. 

\begin{figure}
\begin{center}
\epsfig{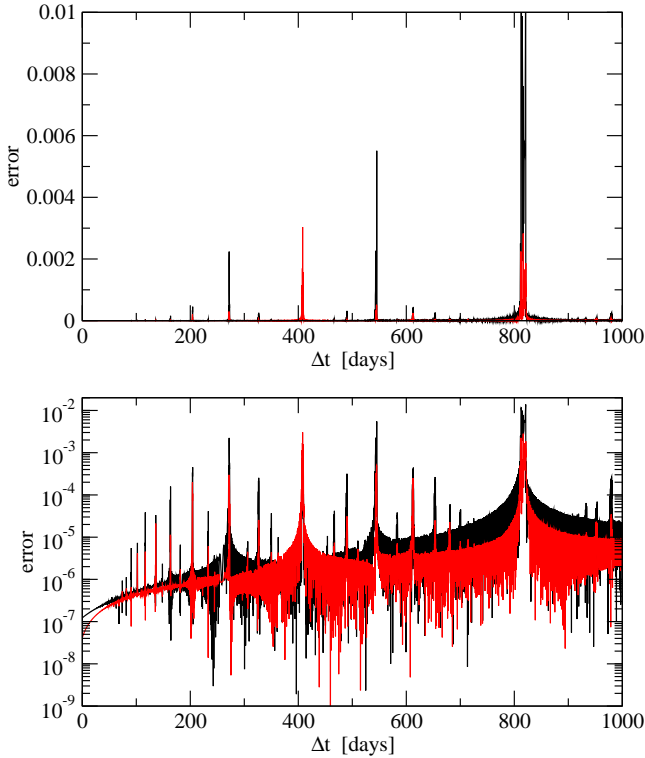}
\caption{Maximum relative errors for the semimajor axis (black) and the eccentricity (red) as function of the step time of MAMA, for a planetesimal of radius $s=5$ km. Top frame shows results in linear scale, while the bottom plot shows the same results in log-scale. Notice the appearance of peaks when the time step $\Delta t$ is commensurate with the orbital period of the planetesimal $T \simeq 820$ days.}
\label{fig7bis}
\end{center}
\end{figure}

Figure \ref{fig7bis} shows the same behavior in more detail, where we compared MAMA and full N-body integrations for a single planetesimal with $a_0=2$ AU and $s=5$ km. Here the time-step interval was extended up to $1000$ days, larger than the orbital period of the particle $T \simeq 820$ days. We can see that the peaks appear precisely at values commensurate with the $T$. The largest occurs at a $1:1$ resonance (i.e. $\Delta t/T = 1/1$), while others are also visible at ratios equal to $2:3$, $1:2$, $1:3$ and $1:4$. This effect seems to be a consequence of the passage from the original ODE to discrete equations which contain an implicit dependence on time in the form of delta functions. 

Since these commensurabilities are an artifact of the algebraic mapping and affect the overall precision of the results, they must be avoided whenever possible. We have therefore adopted a value equal to $\Delta t=150$ days. Although some peaks are still visible in this range (see 
Figure \ref{fig7}), their amplitudes are not very significant and the maximum error in both eccentricity and semimajor axis always seem to be below $10^{-4}$, even for planetesimals with very small radii. It should be stressed, however, that this stepsize is recommended for a TBS  with masses and orbital elements corresponding to $\gamma$-Cephei. If MAMA were to be applied to another binary system, similar tests as those described here should be performed to estimate the best time-step.

\begin{figure}
\begin{center}
\epsfig{figure=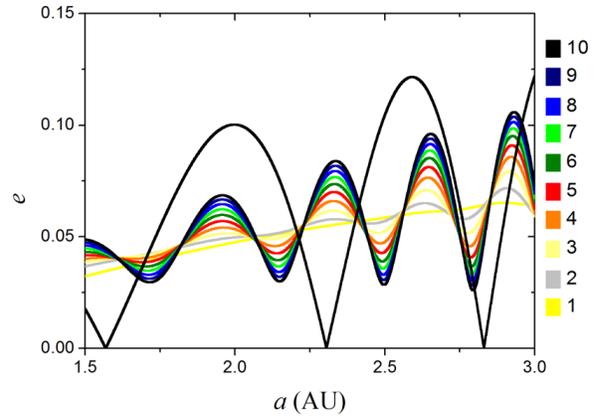,width=8cm,angle=0,clip=}
\caption{Eccentricity as function of the semimajor axis, at $t=100$ binary orbits, of an initial swarm of $10^5$ the particles. The broad black line shows results of the conservative secular model, while other curves correspond to simulations with a non-linear gas drag. All orbital configurations were evolved using MAMA and show excellent agreement with a full N-body simulation.}
\label{fig:paard_001}
\end{center}
\end{figure}

\subsection{Sample Test}

To test MAMA under these conditions, we considered a total sample of $10^5$ initial conditions in circular orbits distributed uniformly between $1$ and $4$ AU, and adopting random values for the mean anomalies. These were separated into $10$ different values of the particle size, between 
$s=1$ and $s=10$ km, considering $10^4$ initial conditions for each radius. Each was then evolved using our map under the gravitational effects of the secondary star of $\gamma$-Cephei plus gas drag. For the gas we assumed an axisymmetric gas disk ($e_{\rm g}=0$, $g_{\rm g}=0$) with the same characteristics as described in Section \ref{sec:nada} ($a_{\rm out}=5$ AU and $\rho_{\rm g} = 5 \times 10^{-10}$ gr/cm$^3$ at $a=2$ AU). 

The results are shown in Figure \ref{fig:paard_001}, were we plot the variation of the eccentricities after 100 binary orbits. Colors identify different particle sizes. For comparison, we also plotted in broad black curves the evolution of the same initial conditions without the effects of gas drag. 

In accordance with the secular equations (\ref{eq6}) and (\ref{eq7}), the particles exhibit an oscillation in eccentricity from $0$ to $2e_{\rm fG}$ with a secular frequency $g_{\rm G}$ that is a function of the semimajor axis. Gas drag causes a systematic damping of the amplitude of oscillation (e.g. Marzari \& School 2000). Thus, the smallest particles show a smaller amplitude of oscillation than their bigger companions. The results obtained with MAMA are in excellent agreement with those presented by other authors (e.g. Th\'ebault et al. 2006, Paardekooper \& Leinhardt 2010).

\subsection{Speed of \textit{MAMA}}\label{sec:velmapa}

Having computed an adequate time-step, and checked its precision, now we turn to the CPU effectiveness of MAMA when compared with a full N-body numerical simulation. Once again we chose the $\gamma$-Cephei system as example, with a $m_{\rm A}$-centric eccentric precessing gas disk with $e_{\rm g}=0.2$ and $g_{\rm g}=-2 \pi / 1000$ 1/yr. For this test we considered a set of 140 planetesimals with radii between $1$ km and $15$ km, which are integrated for a total timespan of $2000$ years.

\begin{figure}
\begin{center}
\centering
\epsfig{figure=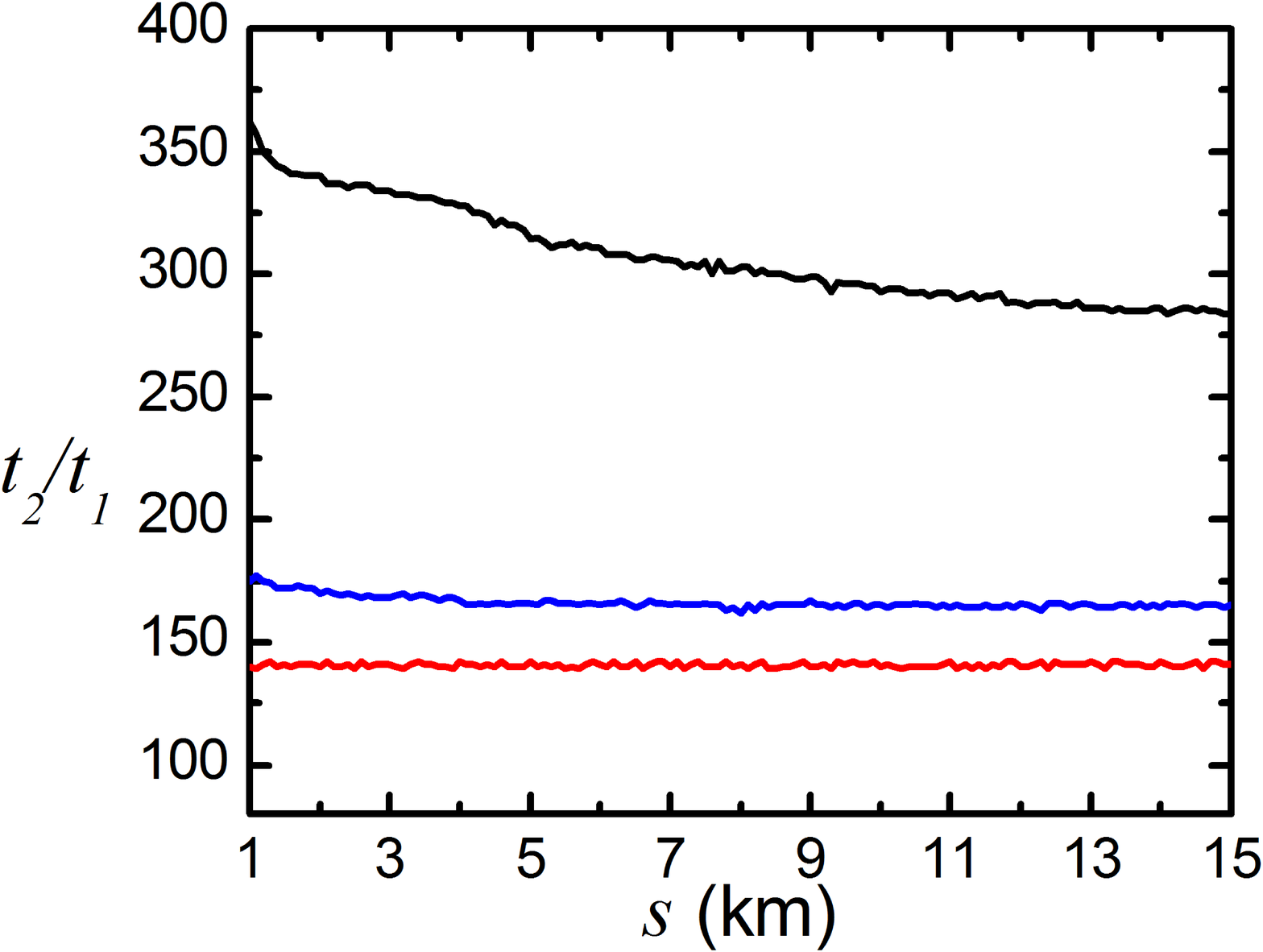,width=8.0cm,angle=0,clip=}
\caption{Ratio of the average computation time between the numerical integrator ($t_2$) and MAMA ($t_1$), for planetesimals of different sizes. The particles start with three different values of the semimajor axis: $1$ AU (black), $2$ AU (blue) and $4$ AU (red).} 
\label{fig:time_comp}
\end{center}
\end{figure}

To solve the exact equations we used a Bulirsch-Stoer integrator with an adaptive step-size, and an error tolerance of $ll=-12$. We carried out three different tests. In all cases initial conditions were chosen with $e=0$, $M=0$, and but with different semimajor axis: $1$ AU, $2$ AU and $4$ AU. We then average the CPU time for the different size-particles and we estimated an averaged time $t_2$ for each set. Finally, we compare these values with those obtained employing MAMA, and denoted these values as $t_1$. The resulting ratio $t_2/t_1$ is shown in Figure \ref{fig:time_comp}, where each  set is plotted using a different color: black ($1$ AU), blue ($2$ AU) and red ($4$ AU). We can see from that MAMA is systematically much faster than the N-body code, although the exact rate depends on the semimajor axis. Even so, the algebraic map is (at worse) 100 times faster than a full integration of the exact equations of motion. 

\begin{figure} 
\begin{center}
\centering
\epsfig{figure=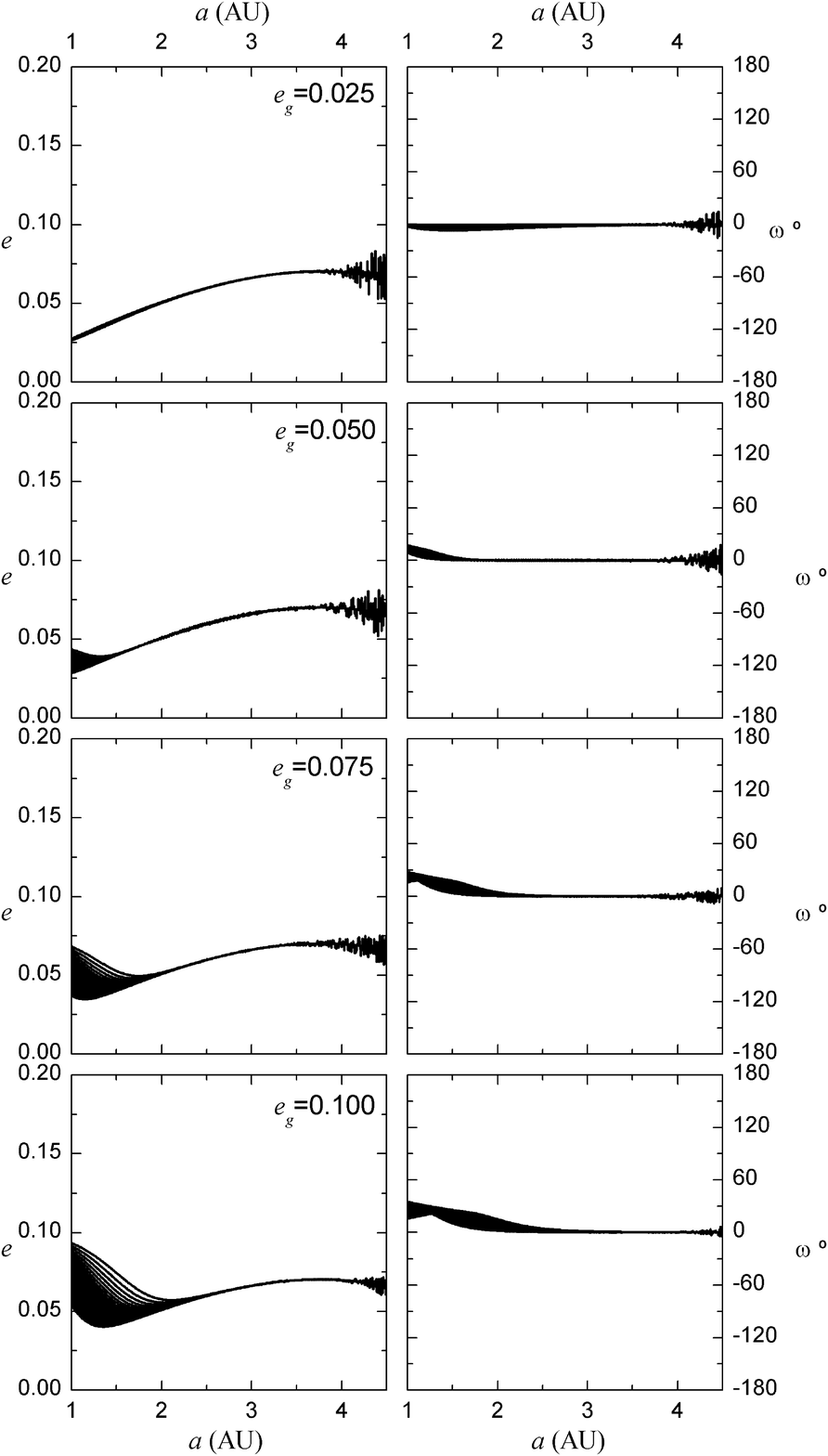,width=8.5cm,angle=0,clip=}
\caption{Eccentricity (left) and longitude of pericenter (right) for planetesimals with sizes between $1km \leq s \leq 10km$ ($\Delta s=0.025km$) as function of the semimajor axis $a$. The graphs show the planetesimals after $3 \times 10^5$ years, when all achieved their equilibrium solutions. For the simulation we assume a static and aligned gas disk ($\omega=0^\circ$, $g_g=0$) and we change its eccentricity: $e_g=0.025$ (top panels), $0.05$, $0.075$ and $0.1$ (bottom panels).}
\label{fig:estatico1}
\end{center}
\end{figure}

\begin{figure} 
\begin{center}
\centering
\epsfig{figure=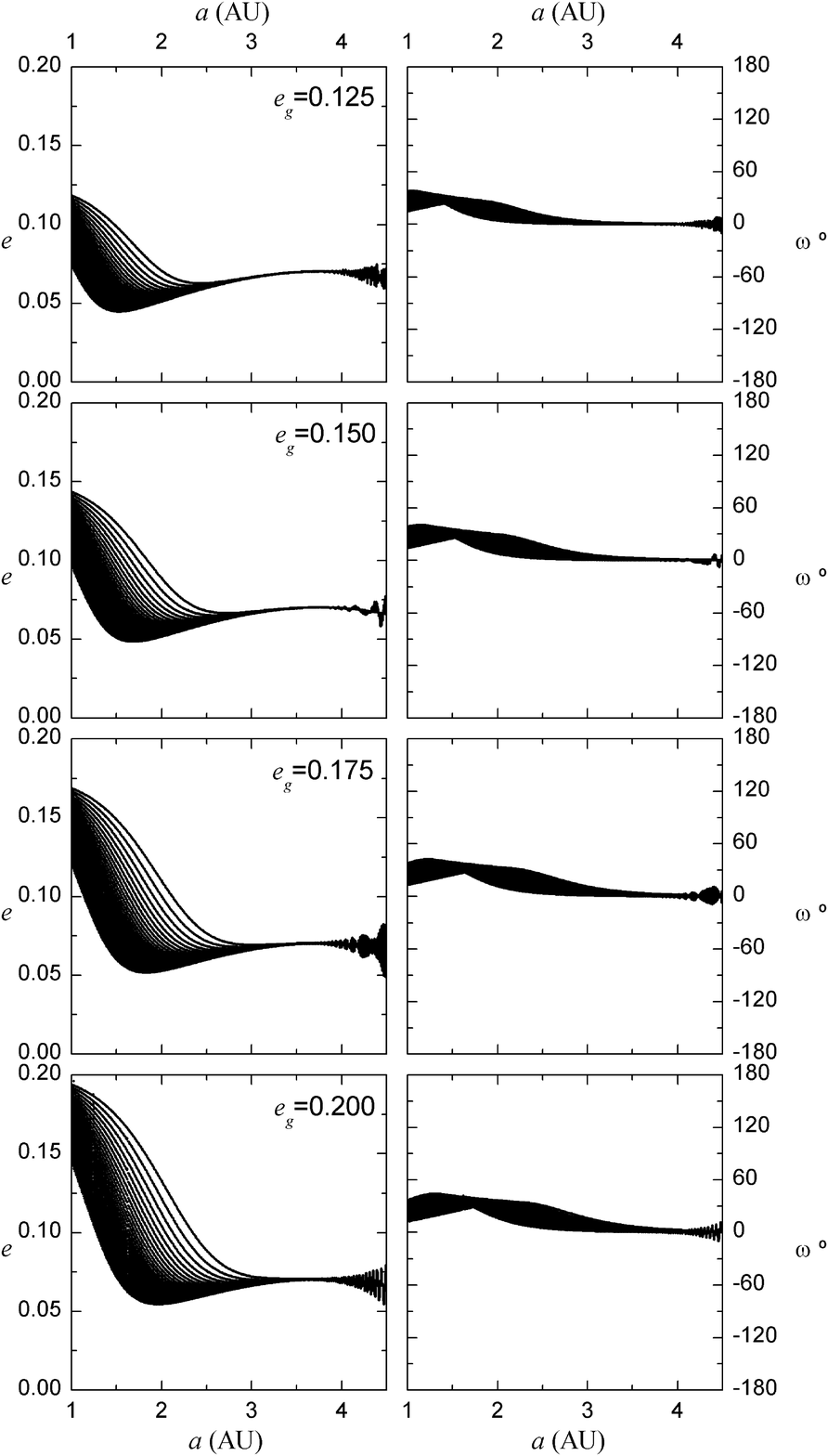,width=8.5cm,angle=0,clip=}
\caption{Eccentricity (left) and longitude of pericenter (right) for planetesimals with sizes between $1km \leq s \leq 10km$ ($\Delta s=0.025km$) as function of the semimajor axis $a$. The graphs show the planetesimals after $3 \times 10^5$ years, when all achieved their equilibrium solutions. For the simulation we assume a static and aligned gas disk ($\omega=0^\circ$, $g_g=0$) and we change its eccentricity: $e_g=0.125$ (top panels), $0.15$, $0.175$ and $0.2$ (bottom panels).} 
\label{fig:estatico2}
\end{center}
\end{figure}

\section{Accretional Conditions in the $\gamma$-Cephei system}\label{sec:buscando}

As discussed in the introduction, the problem of planetesimal accretion in TBS is extremely complex. Although in part this is due to uncertainties in the structure and dynamics of their primordial gaseous disks, it is also affected by our lack of knowledge of the behavior of planetesimal swarms under different disk structures. While the first of these problems are beyond the scope of this work, we may employ MAMA as a working bench to attempt to gain insight on how different disk scenarios may affect collisional velocities and possible accretion among small-size planetesimals.

With these considerations in mind, and for a sake of simplicity, in this section we consider a static (non-precessing) disk with its pericenter aligned with the pericenter of the orbit of the secondary star ($\varpi_{\rm g}=\varpi_{\rm B}=0$). Then, the only free parameter we need to consider is the eccentricity ($e_{\rm g}$) of the disk. We could have chosen to examine the role of any other parameter, but the ellipticity of the gas component is probably the most sensitive one affecting the orbital evolution of small solid bodies.

We analyzed the role of $e_{\rm g}$ considering fixed values between $0.025$ and $0.2$ with steps of $\Delta e_{\rm g} = 0.025$. For each value we generated a total of $16000$ initial conditions for the planetesimals. Their initial semimajor axis were chosen in the interval $1 - 4$ AU and their radii between $1$ and $10$ km. This defined a grid in the $(a,s)$ plane with spacing $\Delta a = 0.1$ AU and $\Delta s = 0.025$ km. All initial orbits were circular with random values of the mean anomaly. 

The secular phase space of planetesimals embedded in a circumstellar disk in a TBS has a stable equilibrium solution. For a static disk, this solution is a fixed point in the $(k,h)$ plane  (Paardekooper et al. 2008, Beaug\'e et al. 2010), while for a precessing disk the stationary orbits are limit cycles (Beaug\'e et al. 2010). The orbital evolution was followed for $3 \times 10^5$ yrs, after which all planetesimals reached their equilibrium solutions. Their final values of the eccentricity $e$ and longitude of pericenter $\varpi$, as function of their initial semimajor axis, are shown in Figures \ref{fig:estatico1} and \ref{fig:estatico2}; the first for disks with $e_{\rm g} \leq 0.1$, while the second presents results for higher values. The different ``curves'' are actually sequences of points made up of planetesimals of equal sizes. While in most cases a size spectrum leads to a significant spread in final values of the secular variables, for values of $e_{\rm g} \sim 0.05$ the solutions appear more 
coherent. 

\begin{figure}
\begin{center}
\centering
\epsfig{figure=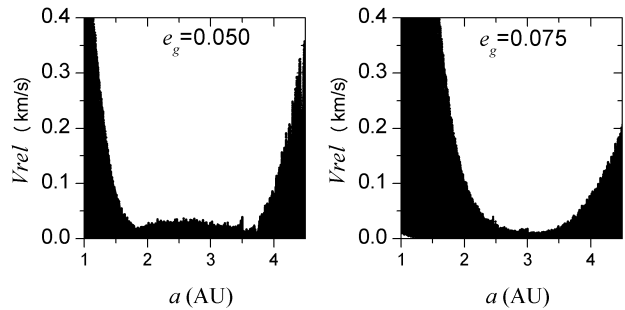,width=8.8cm,angle=0,clip=}
\caption{Distribution of relative velocities $V_{\rm rel}$ for planetesimals of different sizes  ($1 \leq s \leq 10$ km), in an aligned static gas disk with eccentricity $e_{\rm g}=0.05$ (left) and $0.075$ (right). The impact speeds were calculated once the planetesimals achieved their equilibrium solutions. Both cases show a region where $V_{\rm rel}$ is below of the critical limit for disruption, here estimated to be $\sim 70$ m/s (Stewart \& Leinhardt 2009).}
\label{fig:avrel}
\end{center}
\end{figure}

An advantage of our map is that it keeps track of the true longitude of all particles, thus allowing for the identification of possible collisions. We then proceeded to calculate the impact velocities between planetesimal pairs for each value of $e_{\rm g}$. The best results are shown in Figure \ref{fig:avrel}, were we plotted the average relative velocities as function of the semimajor axis. The left plot, corresponding to a disk with $e_{\rm g} \sim 0.05$ shows the most promising scenario, in which most collisions between particles with $a \in [2,3.7]$ AU led to values below $50$ m/s. A slightly higher eccentricity for disk, however, leads to a much smaller accretion-friendly region, now restricted to values close to $3$ AU. 

From Stewart \& Leinhardt (2009) we can estimate the maximum relative velocity before disruption as function of the physical radii of the planetesimals. We found that the worst case scenario occurs for pairs with radii between $1$ and $2$ km, leading to disruption speeds higher than $70$ m/s. Thus, it appears that both examples shown in Figure \ref{fig:avrel} may in fact lead to constructive collisions and serve as breeding grounds for more massive embryos. 

\section{Conclusions}

In this paper we presented an algebraic map, dubbed MAMA, for the dynamical evolution of massless particles embedded in a gas disk, orbiting a central star and perturbed by a secondary stellar component with high eccentricity. Only coplanar motion is considered. The MAMA was constructed combining two models, one for the secular dynamics generated by the gravitational perturbations from the secondary star (Heppenheimer 1978, Th\'ebault et al. 2006, Giuppone et al. 2011), plus a second set of equations modeling the effects of a non-lineal gas drag (Adachi et al. 1976, Weidenschilling et al. 1997, Supulver \& Lin 2000, Beaug\'e et al. 2010).

The map was shown to be precise and able to reproduce the secular dynamics of small planetesimals in circumstellar orbits in tight binary systems (TBS). It is at least 100 times faster than  conventional N-body codes, making it a good work bench with which to study possible scenarios for planetary accretion. 

Although resonant interactions in the outer parts of the gas disk could invalidate the secular approximation, we found that capture can only occur for very small planetesimals. For all other bodies, or semimajor axis below $\sim 3$ AU, the secular model should be fairly precise. 

As an example, we applied MAMA to the $\gamma$-Cephei system, a TBS with a giant exoplanet orbiting its main star at approximately $2$ AU. We analyzed the evolution of $16000$ collision-less particles with sizes between $0.025$ and $20$ km, and distributed from $1$ to $4$ AU. We considered an eccentric and static disk aligned with the orbit of the binary. The eccentricity of the disk was chosen as the test parameter, varying its magnitude between $0.025$ and $0.2$. For each value we simulated the evolution of 16000 particles for $3 \times 10^5$ years, with low computational cost. 

We were able to calculate the relative velocity between all the pairs of particles, and estimate their collisional dynamics. We found that a disk with $e_{\rm g} = 0.05$ appears to define a relatively large region in the semimajor axis domain were impact velocities are sufficiently low to lead to accretion. This region contains the present location of the exoplanet.

Notwithstanding this encouraging result, the aim of this paper was not a detailed and extensive search for accretional scenarios in TBS, but to present a series of examples of possible applications. Future implementations will show whether this or other scenarios may hold the key to planetary formation in these complex systems. 

\section*{Acknowledgments} 
This work was partially financed by the Argentinian Research Council -CONICET- and the Universidad Nacional de C\'ordoba -UNC-.

\end{document}